# Lower bound on the thickness of broadband dielectric mirrors

Mikhail A. Kats

Department of Electrical and Computer Engineering, University of Wisconsin-Madison

mkats@wisc.edu

## Abstract

We present a lower bound on the total physical thickness of a lossless, non-dispersive dielectric mirror for a given minimum reflectance across a wavelength band. The bound is based on a causality-based sum rule translated from a known result in acoustics, that expresses the wavelength integral of the logarithmic transmission of any lossless one-dimensional refractive-index profile as a function of the total thickness of each material comprising the profile, independent of how the layers are arranged. Calculating the thickness bound for a dielectric mirror, given some incident medium and substrate, requires only the lowest and highest refractive indices used in the thin-film stack and the minimum desired reflectance for a given wavelength span. We find that (1) the bound is set by the wavelength span so, for example, a 400–700 nm and an 800–1100 nm mirror have the same bound; (2) the figure of merit for materials minimizing the necessary thickness is $(n_H - n_L)^2/(n_H + n_L)$, where $n_H$ and $n_L$ are the indices of the high- and low-index materials; and (3) each additional "nine" of reflectance (e.g., from 99% to 99.9%) requirement adds a fixed amount of additional thickness to the bound. Non-exhaustive numerical calculations show that the readily achievable thickness is about twice our bound.

## 1. Introduction

Compared to metal mirrors, which have reflectance (R) values in the 90–98% range in the visible and near infrared, multilayer dielectric mirrors routinely exceed $R = 99\%$ and "super-mirrors" can reach 99.999% [1]. However, a dielectric mirror typically only has high performance over a moderate wavelength band, with the reflectance resulting from thin-film interference manifesting as photonic band gaps. Dielectric mirrors are generally designed by thin-film optimization [2, 3, 4, 5], fabricated using thin-film deposition techniques, and commercially available from many vendors, with wider reflectance bands requiring many more layers and more total thickness. It would be useful to have a general lower bound on the thickness of a dielectric mirror for a given bandwidth and reflectance across that bandwidth, to know whether an optimized design can be substantially improved, or whether using different materials would be substantially helpful to make the design thinner. A thematically similar result is the Rozanov bound for metal-backed radar absorbers [6], which relates the thickness of an absorbing stack on a metal backing to how well it can suppress reflection across a wavelength band.

Here, we obtain a lower bound on the physical thickness of a thin-film dielectric mirror given a worst-case reflectance across a particular wavelength band, assuming non-dispersive and lossless dielectric films. The resulting bound needs the following inputs: the lowest and highest refractive indices of the films, the indices of the substrate and superstrate, and the minimum reflectance target over a given wavelength span.

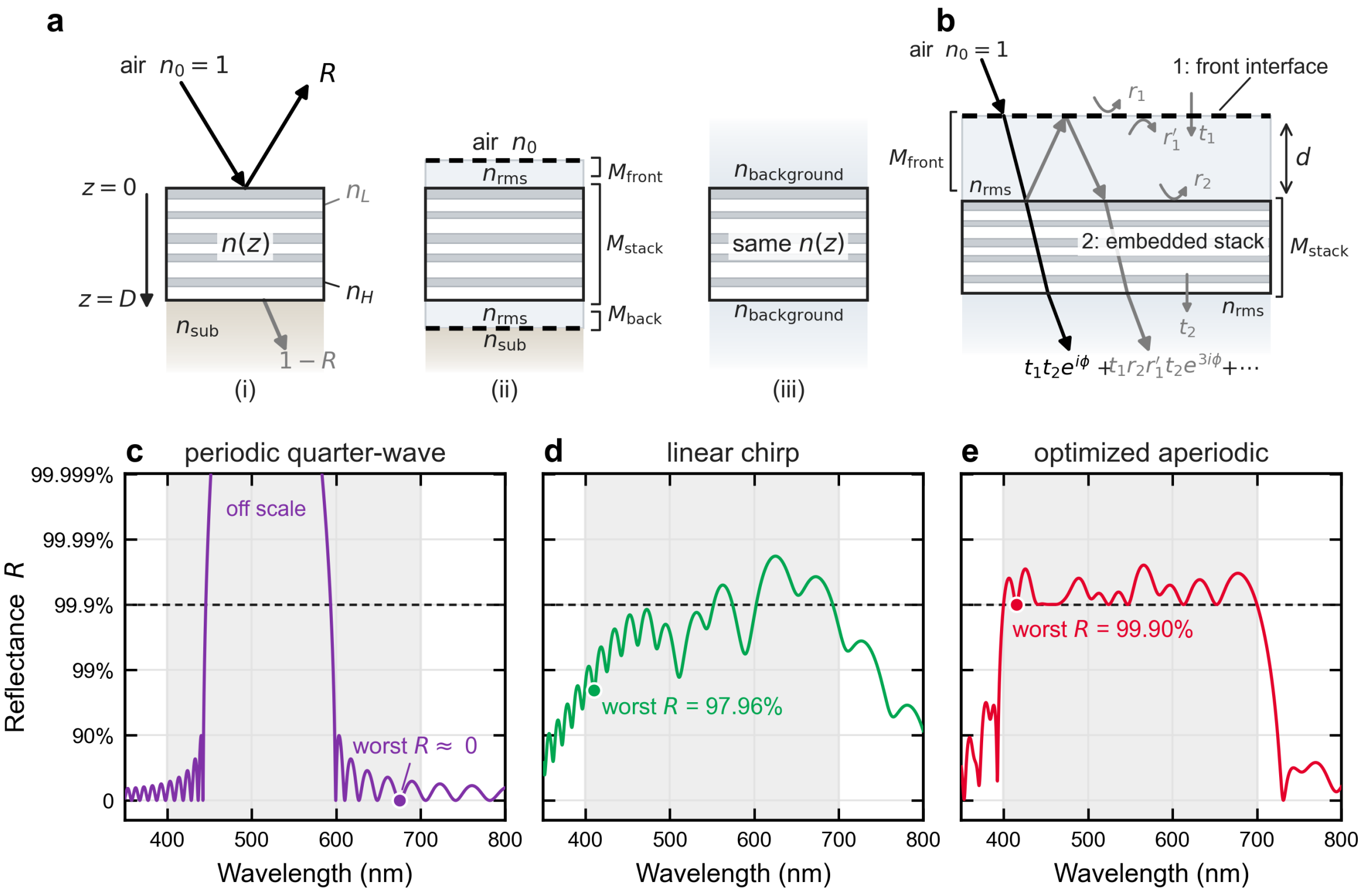

**Figure 1.** (a, b) The configurations used in the derivation, and (c-e) the reflectance $R$ of three types of dielectric mirrors. The representative designs in (c)–(e) aim for at least $R = 99.9\%$ at every wavelength from 400 to 700 nm, at normal incidence, for a mirror thickness $D$ of approximately 3.5 µm, but only one of the designs achieves this target. **(a)** (i): a depth-dependent thin-film assembly with index profile $n(z)$ between depth $z = 0$ and $z = D$, taking values between $n_L$ and $n_H$. The incident medium has index $n_0$ (1 for air), the substrate has index $n_{sub}$, and a large fraction $R$ of the incident power is reflected while $T = 1 - R$ is transmitted. This is the physical mirror whose thickness bound we seek (Section 2.2). (ii): the same profile $n(z)$ with two surrounding layers of index $n_{rms}$, where $n_{rms}$ is the root-mean-square index of the profile. In our derivation, we will take these layers to have zero thickness. The three components making up the structure are labeled by their transfer matrices $M_{front}$, $M_{stack}$, and $M_{back}$. This three-factor decomposition is used in Section 2.2. (iii): the same profile $n(z)$ embedded in a uniform medium of index $n_{background}$ on both sides. The sum rule of Eq. (1) applies to this simplified configuration, which is where the derivation begins (Section 2.1). **(b)** A close-up of the top part of the dielectric mirror, with the $n_{rms}$ layer of thickness $d \to 0$. **(c)–(e)** Reflectance of three designs at essentially the same total thickness (approximately 3.5 µm), with an index pair $n_L = 1.45$, $n_H = 2.30$, roughly corresponding to $SiO_2$/$Nb_2O_5$. The gray band marks the 400–700 nm target band, and the dashed line marks the reflectance target $R = 99.9\%$. **(c)** A periodic quarter-wave stack, which has a stopband narrower than the target band. **(d)** A linearly chirped quarter-wave stack, which gets close, but reaches only $R = 97.96\%$ at its worst point. **(e)** An optimized aperiodic stack, which reaches $R = 99.90\%$ at its worst point.

## 2. Derivation of the bound

### 2.1 Mirror-strength budget and bound for a mirror with symmetric substrate and superstrate

The goal of this paper is to investigate a thin-film stack with non-dispersive index profile $n(z)$ occupying $0 \leq z \leq D$ (i.e., thickness $D$) and taking values between a lowest index $n_L$ and a highest index $n_H$, between an incident medium (i.e., superstrate) of index $n_0$ and a substrate of index $n_{\text{sub}}$ [Fig. 1(a)(i)].

Before examining the situation where the incident medium and substrate are different, we first consider a simpler case, where $n_0 = n_{\text{sub}} = n_{\text{background}}$ [Fig. 1(a)(iii)], for which an exact result is available from one-dimensional acoustics: Norris considered a slab with arbitrarily varying density and compressibility over a finite region in a uniform background, and showed from causality that the frequency integral of its transmission loss is given by a spatial average involving the slab's distribution of local impedance and speed of sound [Eq. (30) of ref. [7]]. That result carries over to normal-incidence electromagnetism with non-magnetic dielectrics as follows.

For any lossless, non-dispersive profile $n(z)$ in a uniform medium of index $n_{\text{background}}$, the integral of $\ln(1/|t|)$ over all wavelengths is fixed by how much material of each index the profile contains, surprisingly irrespective of the order in which the layers are stacked [7]:

$$\int_0^{\infty} \ln \frac{1}{|t(\lambda_0)|} \, \mathrm{d}\lambda_0 = \frac{\pi^2}{2n_{\text{background}}} \int_0^D \left(n(z) - n_{\text{background}}\right)^2 \, \mathrm{d}z. \tag{1}$$

Here $\lambda_0$ is the free-space wavelength and $t$ the complex amplitude transmission coefficient through the entire stack. Note that in this configuration with the same substrate and superstrate, $T = |t|^2$ is the power transmittance. Equation (1) is also the layered-profile case of a sum rule that was derived as an inequality for transmission through periodic screens [8], with the case of a homogeneous slab given explicitly in [9]. Sum rules of this kind follow from causality and passivity [10, 11, 12, 13].

The left side of Eq. (1) adds up the coating's attenuation of transmitted light over all wavelengths, with equal weight per unit wavelength. The right side depends only on how much material of what index is present, and does not depend on the arrangement, so it can be viewed as a fixed "budget" that describes the strength of the mirror [the formal concept of mirror strength is introduced below, in Eq. (2)]. With this viewpoint, the specific design of the coating profile, $n(z)$, is a decision about where in the spectrum this fixed budget is spent.

As an example, Fig. 2(a) shows how the budget of a periodic quarter-wave stack (i.e., a Bragg mirror) is distributed: a large spike at the fundamental stopband centered at wavelength $\lambda_c$, and narrower spikes at the harmonics $\lambda_c/3$, $\lambda_c/5$, etc., with a substantial share (about 19%) of the budget spent at wavelengths shorter than the fundamental reflectance band. In fact, a coating built from two discrete materials cannot avoid distributing some of the budget outside of the main stopband, because the square-wave-like profile of $n(z)$ has harmonics besides the fundamental frequency [4]. A smoothly graded profile can suppress the harmonics, known as a rugate coating [14], though we do not explicitly investigate graded coatings here. In Fig. 2(b), we checked Eq. (1) numerically using the transfer-matrix method, testing several types of thin-film assemblies with $n_0 = n_{\text{sub}} = n_{\text{background}}$ taking values from 1 to 2, confirming the relationship.

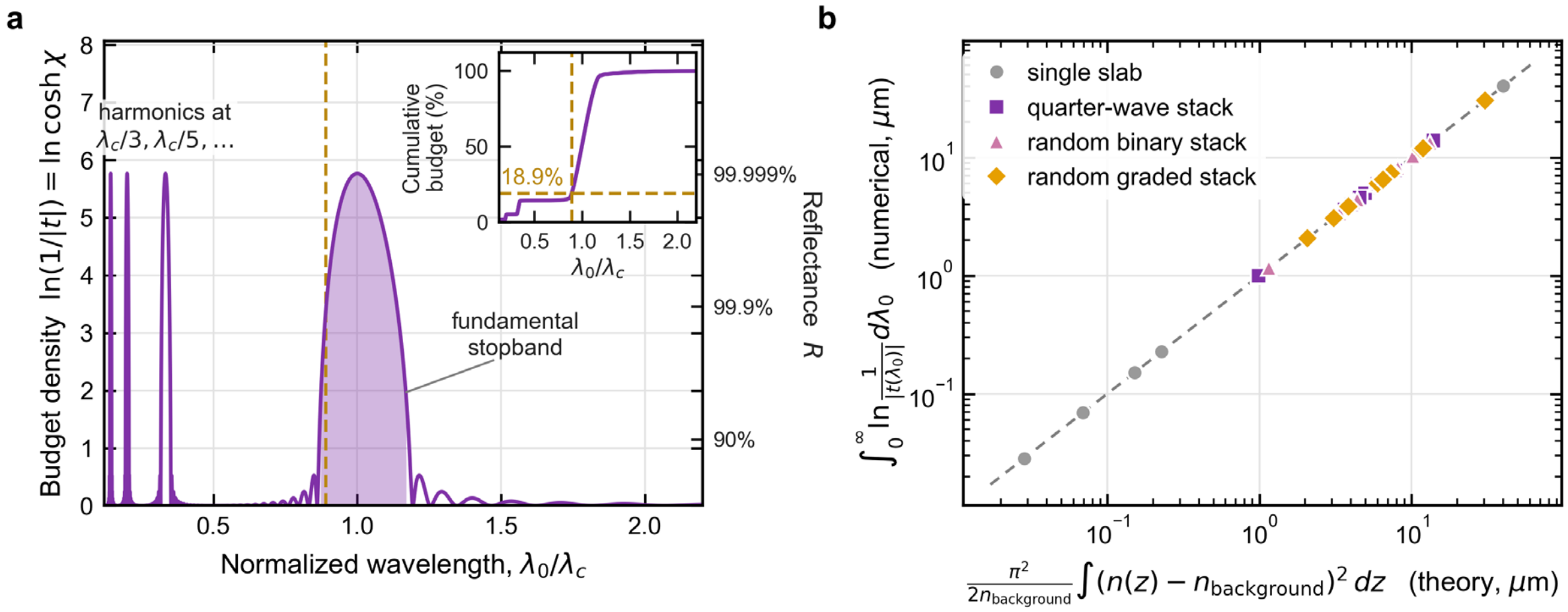

**Figure 2. (a)** The integrand $\ln(1/|t|)$ from Eq. (1), interpreted as a mirror-strength "budget" density, plotted vs. wavelength normalized to the center wavelength $\lambda_c$, for a periodic quarter-wave stack with the 1.45/2.30 index pair, with the same substrate and superstrate. The right axis gives the corresponding reflectance. The inset shows the cumulative budget, where the dashed line marks $100(1 - 8/\pi^2) = 18.9\%$, the share of the budget spent on wavelengths shorter than the fundamental stopband of the quarter-wave stack; the vertical dashed line marks the wavelength at which the cumulative budget crosses this share. **(b)** Eq. (1) tested using a variety of thin-film assemblies: single slabs, quarter-wave stacks, random binary stacks (alternating 1.45/2.30-index layers with random layer thicknesses), and random graded stacks (a random index and a random thickness for every layer), each in a background medium with a randomly drawn index, calculated via the transfer-matrix method. The horizontal axis is the right-hand side of Eq. (1) and the vertical axis is the left-hand side.

Eq. (1) is in terms of transmittance, but we are more naturally interested in reflectance. Reflectance itself is an awkward variable here because, for good mirrors, every value of interest crowds just below $R = 1$. We therefore define a logarithmic measure that we call the mirror strength $\chi$, that quantifies how strongly a structure reflects and is simply related to the integrand of Eq. (1):

$$\chi = \operatorname{atanh}\sqrt{R} \approx \frac{1}{2}\ln\frac{4}{1-R}. \tag{2}$$

For a lossless coating with the same substrate and superstrate, the transmitted power is $|t|^2 = 1 - R$, so $1/|t| = \cosh\chi$ and the integrand of Eq. (1) is $\ln(\cosh\chi)$. For example, $\chi$ is 2.99 at $R = 99\%$, 4.15 at 99.9%, 5.30 at 99.99%, rising by $\frac{1}{2}\ln 10 = 1.151$ for each additional "nine" of reflectance. For a uniform Bragg grating, the mirror strength at the design wavelength is

equivalent to the grating strength $\kappa L$ from coupled-mode theory [15], where $\kappa$ is the coupling coefficient and $L$ the grating length. A requirement of minimum reflectance $R_{\text{target}}$ across a band of free-space wavelengths, from $\lambda_0 = \lambda_{\min}$ to $\lambda_{\max}$, is then the requirement that $\chi \geq \chi_{\text{target}}$ at every wavelength in the band, where

$$\chi_{\text{target}} = \operatorname{atanh}\sqrt{R_{\text{target}}}, \tag{3}$$

and $R_{\text{target}}$ is the specified minimum reflectance. The budget of Eq. (1) limits how much mirror strength the coating of a particular thickness and index values can supply.

### 2.2 Bound extended to dielectric mirrors with non-symmetric substrate and superstrate

Equation (1) is a statement about the thin-film stack embedded in some uniform medium of index $n_{\text{background}}$ (i.e., the symmetric case, where $n_0 = n_{\text{sub}} = n_{\text{background}}$). To convert this into a bound for mirrors with different substrates and superstrates, we go through a two-step process where we first find which $n_{\text{background}}$ provides the minimum budget for the symmetric case, and then use that to create the bound for the asymmetric case [Fig. 1(a)(i)].

In Eq. (1), both $t$ and the right-hand side change with $n_{\text{background}}$, so each choice gives a slightly different sum rule. Each of these sum rules is an exact statement about that particular configuration of thin-film assembly and substrate and superstrate. Any of these sum rules can be carried through the second step (below) to give a bound for the mirror with different substrate and superstrate, so $n_{\text{background}}$ can be chosen to find the tightest bound for the realistic, asymmetric case.

In the following, we write the right-hand side of Eq. (1) in terms of the root-mean-square and mean indices of the profile, making its dependence on $n_{\text{background}}$ explicit and easy to minimize. First, we calculate the root-mean-square index of the profile:

$$n_{\text{rms}}^2 = \frac{1}{D}\int_0^D n(z)^2\,\mathrm{d}z = f n_H^2 + (1-f) n_L^2, \tag{4}$$

where the right-hand side is for a two-material thin-film coating in which the volume (or, equivalently, thickness) fill fraction $f$ is for the high-index material. We also calculate the mean index: $n_{\text{mean}} = D^{-1}\int_0^D n(z)\,\mathrm{d}z$. Then, the right-hand side of Eq. (1) becomes

$$\frac{\pi^2}{2n_{\text{background}}}\left(n_{\text{rms}}^2 - 2n_{\text{background}}n_{\text{mean}} + n_{\text{background}}^2\right)D. \tag{5}$$

Setting the derivative of Eq. (5) with respect to $n_{\text{background}}$ to zero yields $n_{\text{background}} = n_{\text{rms}}$, meaning that the situation with the smallest mirror-strength budget given a symmetric substrate and superstrate is for the stack to be embedded in $n_{\text{rms}}$. Substituting the result back into Eq. (1) gives

$$\int_0^\infty \ln\frac{1}{|t(\lambda_0)|}\,\mathrm{d}\lambda_0 = \pi^2(n_{\mathrm{rms}}-n_{\mathrm{mean}})D. \tag{6}$$

The actual mirror we aim to study has superstrate index $n_0$ and substrate index $n_{\mathrm{sub}}$, which generally differ from each other and from $n_{\mathrm{rms}}$. We relate the real case to the symmetric results by adding two index steps: one from $n_0$ to $n_{\mathrm{rms}}$ at the front face, and one from $n_{\mathrm{rms}}$ to $n_{\mathrm{sub}}$ at the back face [Fig. 1(a)(ii)]. Writing $M_{\mathrm{front}}$ for the transfer matrix of the step from $n_0$ to $n_{\mathrm{rms}}$ and propagation through $n_{\mathrm{rms}}$ (which will be negligible due to vanishing thickness), $M_{\mathrm{stack}}$ for the stack embedded in $n_{\mathrm{rms}}$, and $M_{\mathrm{back}}$ for the propagation through $n_{\mathrm{rms}}$ and the step from $n_{\mathrm{rms}}$ to $n_{\mathrm{sub}}$, the product of the transfer matrices is the transfer matrix of the physical mirror:

$$M_{\mathrm{real}} = M_{\mathrm{back}}\, M_{\mathrm{stack}}\, M_{\mathrm{front}}. \tag{7}$$

Note that in our notation, the interfaces from $n_{\mathrm{rms}}$ to the outermost layers of the embedded stack are included in $M_{\mathrm{stack}}$.

To show how the factors in Eq. (7) combine, we can look first at the first part, $M_{\mathrm{stack}}M_{\mathrm{front}}$. The $n_{\mathrm{rms}}$ region between the step and the stack (which we call the spacer) has vanishing thickness $d \to 0$ [Fig. 1(b)]. Call the front interface "section 1" and the embedded stack including its interfaces with $n_{\mathrm{rms}}$ "section 2", with the spacer between the two sections. Let $t_1$ and $t_2$ be the amplitude transmission coefficients of the two sections alone, $r_1$ and $r_2$ their amplitude reflection coefficients for light arriving from the front, and $r_1'$ the reflection coefficient of section 1 for light arriving from its back side.

These coefficients, used from here through Eq. (10), are flux-normalized: each is defined so that its squared modulus is the fraction of the power transmitted or reflected, so $|r_i|^2 + |t_i|^2 = 1$ for each lossless section. For a region of thin films with the same medium on both sides, these are the usual field-amplitude ratios of the reflected or transmitted to the incident field. Section 1, however, has $n_0$ in front and $n_{\mathrm{rms}}$ behind, and there the usual ratio of transmitted to incident field amplitude must be multiplied by $\sqrt{n_{\mathrm{out}}/n_{\mathrm{in}}}$, where $n_{\mathrm{in}}$ and $n_{\mathrm{out}}$ are the indices of the entry and exit media, for its square to be the transmitted power fraction [16]. The reflection coefficient needs no factor because the incident and reflected waves are in the same medium. When the coefficients of adjacent sections are multiplied, the factors of each shared intermediate medium cancel in pairs, so flux-normalized coefficients combine by the same rules as the ordinary ones, and the transfer matrices in Eq. (7) act on field amplitudes as usual.

Light can cross both sections directly, or be reflected back and forth across the (actually vanishingly small) spacer any number of times before leaving [Fig. 1(b)]. Each crossing of the $n_{\mathrm{rms}}$ region adds the propagation phase $\phi = 2\pi n_{\mathrm{rms}} d/\lambda_0 \to 0$, and each additional round trip multiplies a path's amplitude by $r_2 r_1' e^{2i\phi} \to r_2 r_1'$. Summing the geometric series describing this multi-path interference gives the amplitude transmission coefficient $t$ of the two sections combined, which is the familiar Airy formula [4], written for general $\phi$ and immediately evaluated in the $\phi \to 0$ limit of the vanishing spacer:

$$t = \frac{t_1 t_2 e^{i\phi}}{1 - r_2 r_1' e^{2i\phi}} = \frac{t_1 t_2}{1 - r_2 r_1'}. \tag{8}$$

The denominator of Eq. (8) has a magnitude of at most $1 + |r_1||r_2|$, since $|r_1'| = |r_1|$ for a section of the stack comprising lossless materials. We can convert the reflection and transmission coefficients of each section $i$ to its mirror strength $\chi_i$ using Eq. (2): $|r_i| = \tanh \chi_i$, so $|t_i| = (1-|r_i|^2)^{1/2} = 1/\cosh \chi_i$, and therefore $1/|t_i| = \cosh \chi_i$ and $|r_i|/|t_i| = \tanh \chi_i \cosh \chi_i = \sinh \chi_i$. Taking the magnitude of Eq. (8) then gives

$$\frac{1}{|t|} \leq \frac{1 + |r_1||r_2|}{|t_1||t_2|} = \cosh \chi_1 \cosh \chi_2 + \sinh \chi_1 \sinh \chi_2 = \cosh(\chi_1 + \chi_2). \tag{9}$$

Thus we find that

$$\chi_{1+2} \leq \chi_1 + \chi_2, \tag{10}$$

where $\chi_{1+2}$ is the mirror strength of the two sections combined. This means that the mirror strength of two lossless sections in series never exceeds the sum of the mirror strengths of the two sections evaluated individually, with equality when the interference phase aligns the two contributions. Joining the third factor of Eq. (7) in the same way shows that the mirror strength of the real mirror is at most the sum of the contributions of its three factors. A single index step from $n_a$ to $n_b$ has Fresnel coefficient $r_{ab} = (n_a - n_b)/(n_a + n_b)$ and hence $\chi_{ab} = \operatorname{atanh}|r_{ab}| = \frac{1}{2}|\ln(n_b/n_a)|$. The two index steps, from $n_0$ to the vanishingly thin $n_{\text{rms}}$ region, and from $n_{\text{rms}}$ to $n_{\text{sub}}$, together contribute at most mirror strength $\chi_{\text{steps}}$ to the total mirror strength $\chi$:

$$\chi_{\text{steps}} = \frac{1}{2}\left|\ln \frac{n_{\text{rms}}}{n_0}\right| + \frac{1}{2}\left|\ln \frac{n_{\text{rms}}}{n_{\text{sub}}}\right|. \tag{11}$$

Our target physical dielectric mirror must satisfy $\chi \geq \chi_{\text{target}}$ at every wavelength in the prescribed wavelength band, and we have now determined that its mirror strength is at most $\chi_{\text{embedded}} + \chi_{\text{steps}}$, where $\chi_{\text{embedded}}$ is the mirror strength of the embedded coating, that is, of the stack surrounded by $n_{\text{rms}}$ on both sides [Fig. 1(a)(ii)]. From the perspective of a design specification, the embedded coating must therefore have $\chi_{\text{embedded}} \geq \chi_{\text{target}} - \chi_{\text{steps}}$ at every wavelength in the target band. If $\chi_{\text{steps}} \geq \chi_{\text{target}}$, this requirement is empty, so from here on we take the difference $\chi_{\text{target}} - \chi_{\text{steps}}$ to be positive, which is anyway the case for any practical mirror target.

The right-hand side of Eq. (6) contains $n_{\text{rms}} - n_{\text{mean}}$, which depends on the index profile of the stack. For every profile bounded by $n_L \leq n(z) \leq n_H$, the product $(n_H - n(z))(n(z) - n_L)$ is non-negative at every depth $z$. Expanding it, $(n_H + n_L)\, n(z) - n(z)^2 - n_H n_L \geq 0$. Averaging over the thickness turns $n(z)^2$ into $n_{\text{rms}}^2$ [Eq. (4)] and $n(z)$ into $n_{\text{mean}}$, so $n_{\text{mean}} \geq (n_{\text{rms}}^2 + n_H n_L)/(n_H + n_L)$, and therefore

$$n_{\mathrm{rms}} - n_{\mathrm{mean}} \leq \frac{(n_H - n_{\mathrm{rms}})(n_{\mathrm{rms}} - n_L)}{n_H + n_L}, \tag{12}$$

with equality when the profile takes only the two values $n_H$ and $n_L$, because then the product $(n_H - n(z))(n(z) - n_L)$ vanishes at every depth and its average is zero. For a two-valued profile, $n_{\mathrm{mean}}(f) = f n_H + (1 - f) n_L$, where $f$ is the volume (or thickness) fraction of the high-index material, as in Eq. (4). Thus, among all profiles that have the same $n_{\mathrm{rms}}$, a two-valued profile has the smallest $n_{\mathrm{mean}}$, and therefore the largest budget $\pi^2(n_{\mathrm{rms}} - n_{\mathrm{mean}})D$ in Eq. (6). Using this largest value gives a thickness bound that holds for every profile with indices between $n_L$ and $n_H$.

From Eq. (6), for the stack embedded in a uniform medium of index $n_{\mathrm{background}} = n_{\mathrm{rms}}$, $\int_0^\infty \ln(1/|t(\lambda_0)|)\, \mathrm{d}\lambda_0 = \pi^2(n_{\mathrm{rms}} - n_{\mathrm{mean}})D$. The integrand is never negative and, for a stack that meets the target specification, at every wavelength inside the target band it is at least $\ln\left(\cosh\left(\chi_{\mathrm{target}} - \chi_{\mathrm{steps}}\right)\right)$, because there $1/|t| = \cosh \chi_{\mathrm{embedded}} \geq \cosh\left(\chi_{\mathrm{target}} - \chi_{\mathrm{steps}}\right)$. The full integral is therefore at least the width of the band, $\Delta\lambda = \lambda_{\max} - \lambda_{\min}$, multiplied by this in-band value, i.e., $\pi^2(n_{\mathrm{rms}} - n_{\mathrm{mean}})D \geq \Delta\lambda \ln\left(\cosh\left(\chi_{\mathrm{target}} - \chi_{\mathrm{steps}}\right)\right)$. Solving for $D$ gives one thickness bound for each fill fraction $f$, and every profile obeys the bound whose $f$ matches its $n_{\mathrm{rms}}$ [Eq. (4)]. The design's $n_{\mathrm{rms}}$ is not known in advance, so we keep the smallest of these bounds, resulting in the general bound:

$$D \geq \Delta\lambda \min_{0<f<1} \frac{\ln\left(\cosh\left(\chi_{\mathrm{target}} - \chi_{\mathrm{steps}}(f)\right)\right)}{\pi^2(n_{\mathrm{rms}}(f) - n_{\mathrm{mean}}(f))}. \tag{13}$$

Here, $\min_{0<f<1}$ denotes the smallest value of the expression to its right over all fill fractions $0 < f < 1$, and $\chi_{\mathrm{steps}}(f)$ is Eq. (11) evaluated at $n_{\mathrm{rms}}(f)$. The minimization in Eq. (13) is over $f$ and can be carried out numerically for any $n_0$ and $n_{\mathrm{sub}}$.

To obtain a closed form of the bound, we again consider the special symmetric case $n_{\mathrm{sub}} = n_0$ and assume $n_0 \leq n_L$, so that $n_{\mathrm{rms}} \geq n_0$; then the two steps of Eq. (11) are equal, and $\chi_{\mathrm{steps}} = \ln(n_{\mathrm{rms}}/n_0)$. Then, two replacements allow the minimization of Eq. (13) to be carried out analytically. First, we replace $\ln(\cosh \chi)$ by its lower bound $(\chi - \ln 2)$, which follows from $\cosh \chi = (e^{\chi} + e^{-\chi})/2 \geq e^{\chi}/2$ and turns out to barely affect the bound. Second, we replace $\ln(n_{\mathrm{rms}})$ in Eq. (11) by its first-order Taylor approximation about $n_{\mathrm{rms}} = (n_H + n_L)/2$; because $\ln(n_{\mathrm{rms}})$ is concave, the tangent line lies above $\ln(n_{\mathrm{rms}})$, so this replacement only makes $\chi_{\mathrm{steps}}$ larger. Setting the derivative of the resulting right-hand side of Eq. (13) with respect to $f$ to zero then gives the closed form of the bound:

$$D \geq \frac{\mathcal{K}\Delta\lambda}{2}\left(\chi_{\mathrm{net}} + \sqrt{\chi_{\mathrm{net}}^2 - r_{HL}^2}\right), \quad \mathcal{K} = \frac{4(n_H + n_L)}{\pi^2(n_H - n_L)^2}, \tag{14}$$

where

$$\chi_{\text{net}} = \chi_{\text{target}} - \chi_{\text{offset}}, \quad \chi_{\text{offset}} = \ln\left(\frac{n_H + n_L}{n_0}\right), \quad r_{HL} = \frac{n_H - n_L}{n_H + n_L}. \tag{15}$$

Here $r_{HL}$ is the Fresnel coefficient of a single step from $n_H$ to $n_L$. The offset has two components, one from each of the two replacements above. The tangent-line replacement bounds $\chi_{\text{steps}}$ from above by its value at the tangent point, $\ln((n_H + n_L)/(2n_0))$, plus a term proportional to $n_{\text{rms}} - (n_H + n_L)/2$, and only the term proportional to $n_{\text{rms}} - (n_H + n_L)/2$ stays inside the minimization. The constant value combines with the $\ln 2$ from the cosh replacement to give the offset: $\ln((n_H + n_L)/(2n_0)) + \ln 2 = \ln((n_H + n_L)/n_0) = \chi_{\text{offset}}$. Thus, $\chi_{\text{net}}$ is the part of the target that remains after these two constants are subtracted. In this symmetric case ($n_{\text{sub}} = n_0$), Eq. (14) is valid for $\chi_{\text{net}} \geq r_{HL}$ and $n_0 \leq n_L$.

From Eq. (11), for the condition $n_0 \leq n_{\text{sub}} \leq n_{\text{rms}}^2/n_0$, the two index steps together contribute no more mirror strength than in the symmetric case $n_{\text{sub}} = n_0$. Equation (14) therefore remains a valid, slightly conservative lower bound. The upper limit $n_{\text{rms}}^2/n_0$ depends on the profile through $n_{\text{rms}}$, but there is a profile-independent sufficient condition, $n_0 \leq n_{\text{sub}} \leq n_L^2/n_0$, which still readily covers the majority of practically important thin-film cases. For example, for the 1.45/2.30 pair in air, this condition allows substrate indices from 1 to 2.10. For substrates outside this range, the bound Eq. (13) should be minimized directly, but this can be done straightforwardly by inputting only $n_0$, $n_{\text{sub}}$, $n_L$, $n_H$, the wavelength span $\Delta\lambda$, and the reflectance target.

To summarize: subject to the relatively trivial condition of $\chi_{\text{target}} - \chi_{\text{steps}} > 0$ assumed above, Eq. (13) minimized numerically over $f$ yields the bound for arbitrary $n_0$ and $n_{\text{sub}}$. We also have an analytical closed-form bound, stated in Eq. (14), which is valid for every profile when $n_0 \leq n_{\text{sub}} \leq n_L^2/n_0$ and for a given profile when $n_0 \leq n_{\text{sub}} \leq n_{\text{rms}}^2/n_0$. The analytical bound of Eq. (14) is slightly looser than the numerical bound of Eq. (13).

As an example, for a 400–700 nm mirror in air on both sides ($n_{\text{sub}} = n_0 = 1$) with $R = 99.9\%$ and the 1.45/2.30 pair, roughly representative of $SiO_2$/$Nb_2O_5$, Eq. (14) gives $D \geq 1.78$ µm. Minimizing Eq. (13) numerically for the same case also gives $D \geq 1.78$ µm. Meanwhile, the optimized design shown in Fig. 1(e) meets the target specification at $D = 3.52$ µm, roughly twice the bound. If the same mirror target is instead specified on a substrate, the bound rises slightly: minimizing Eq. (13) numerically gives a bound of $D \geq 1.91$ µm for a substrate of BK7 glass ($n_{\text{sub}} = 1.52$) and $D \geq 1.96$ µm for sapphire ($n_{\text{sub}} = 1.77$). Equation (14) does not include $n_{\text{sub}}$, so it cannot capture this increase. We did not find a closed form analogous to Eq. (14) for the asymmetric case.

From the form of Eq. (14), we obtain the three consequences stated in the abstract, so far assuming the materials are lossless and non-dispersive, and assuming normal incidence: (1) for fixed indices and reflectance target, the minimum-thickness bound is directly proportional to the wavelength span $\Delta\lambda$; (2) at high reflectance, the material figure of merit for minimizing thickness is $(n_H - n_L)^2/(n_H + n_L)$; and (3) each additional "nine" of reflectance, meaning another factor of ten in $1 - R$, adds a constant amount of thickness ($1.15\mathcal{K}\Delta\lambda$) to the bound.

## 3. Numerical tests

To numerically test the bounds, we searched for thin-film stacks meeting each of more than 80 different mirror target specifications with $n_{\text{sub}} = n_0 = 1$, and compared the thinnest design found for each target with Eq. (13). The target specifications we selected cover wavelength spans $\Delta\lambda$ from 60 to 800 nm for bands starting at 400 nm and starting at 800 nm, worst-case reflectance in these bands from $R = 99\%$ to $R = 99.999\%$, and twelve different index pairs with non-dispersive and lossless $n_L$ from 1.38 to 1.70 and $n_H$ from 1.90 to 2.40, all at normal incidence. Figure 3 shows three cross-sections through this set of optimized results, as a function of $\Delta\lambda$, worst-case reflectance target, and as a function of the material figure of merit $(n_H - n_L)^2/(n_H + n_L)$.

Every design was produced with the same optimizer, which, at a fixed total thickness, repeatedly adjusts all the layer thicknesses to attempt to raise the worst-case mirror strength $\chi$ across the target wavelength band. Each proposed adjustment was computed from the derivatives of the mirror strength with respect to the layer thicknesses (i.e., sequential linear programming [17]), and was kept only if recomputing the spectrum showed that the true worst-case reflectance (i.e., lowest reflectance anywhere in the band) improved. The procedure is essentially hill climbing on the worst-case mirror strength. Because such hill climbs stop at local optima, we restarted the search at each thickness from several chirped, periodic, and random starting profiles, and the best result was then perturbed at random dozens of times, keeping each perturbation only when it led to an improvement. The smallest total thickness at which the target is reached was found by bisection, i.e., repeatedly testing the midpoint between a thickness that we found to reach the target and one that did not. For each target specification, the entire procedure was repeated with up to nine pseudorandom seeds and the thinnest design was recorded.

Figure 3(a) shows the thinnest designs found for the 1.45/2.30 pair at $R = 99.9\%$ as the target band is widened, for bands starting at 400 nm and at 800 nm. For all but the narrowest target bands, the optimized designs settle to about 1.8–2.0 times the thickness bound.

Here and throughout the paper, we consistently find that the optimized designs land at roughly 2× the thickness bound. Much of this factor likely comes from the fact that the bound allows the entire integral in Eq. (1) to be "spent" inside the target band, whereas a stack of two discrete materials cannot "spend" this budget so precisely. Some reflection is also produced at the shorter-wavelength harmonics of the layer structure [as in Fig. 2(a)] or just outside the band edges, since it is challenging to make a very sharp transition from highly reflecting to not, as a function of wavelength. Furthermore, the ideal design that gets close to the bound would have a reflectance that exactly matches the target; instead, in the optimized designs, some of the reflectance is slightly above the target [e.g., Fig. 1(e)], thus spending down some of the budget. For example, for the optimized design of Fig. 1(e), splitting the integral in Eq. (1) by wavelength shows that 64% of the budget is spent inside the 400–700 nm target band, while only 51% is needed to exactly meet the target; 23% is spent at shorter wavelengths (including the harmonics), and 13% at longer wavelengths.

At the shortest wavelength span $\Delta\lambda$, the thickness obtained using the optimizer is about 3 to 5 times the bound at the narrowest bands tested [Fig. 3(a)]. This reflects a change of regime: when the fractional bandwidth $B = 2(\lambda_{\max}-\lambda_{\min})/(\lambda_{\max}+\lambda_{\min})$ is below the natural fractional stopband width of a quarter-wave stack, $B_\infty = (4/\pi)\operatorname{asin}\big((n_H-n_L)/(n_H+n_L)\big)$ [16], a single quarter-wave stopband can cover the entire band, and the thickness required is set by the stopband physics rather than by the sum rule. The bound remains valid throughout this regime, but it is far from being a tight bound.

In Fig. 3(b), we varied the minimum reflectance target from $R = 99\%$ to $R = 99.999\%$ for the 400–700 nm wavelength band. Each additional "nine" is found to add a nearly constant amount to the bound (approximately 0.73 µm), and the trend is the same for the optimized designs, which require approximately 1.2 µm in thickness per "nine". In Fig. 3(c), we fixed both the wavelength band and minimum reflectance, but the low ($n_L$) and high ($n_H$) indices are varied. The thickness is plotted vs. the material figure of merit $(n_H - n_L)^2/(n_H + n_L)$ from our bound, and the optimum values are again roughly 2× the bound.

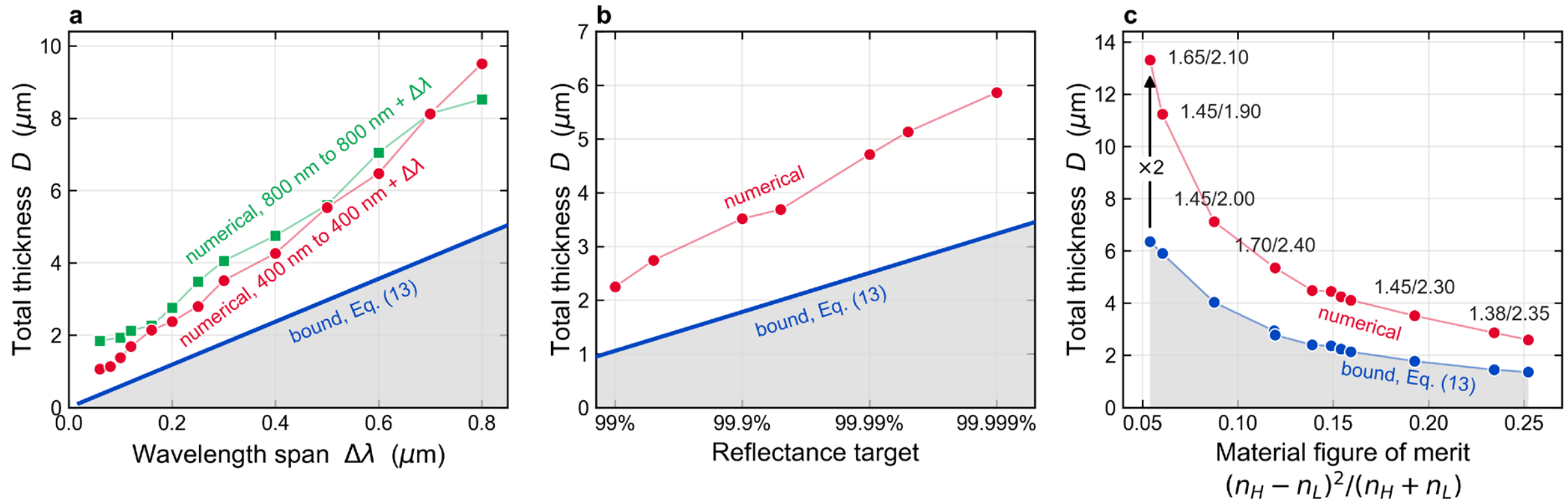

**Figure 3.** The thinnest optimized dielectric-mirror designs, with $n_{\text{sub}} = n_0 = 1$ (symbols; the lines are just a guide for the eye), compared with Eq. (13). **(a)** Total thickness vs. wavelength span for the $R = 99.9\%$ target using the 1.45/2.30 index pair, for wavelength bands starting at 400 nm and at 800 nm. **(b)** Total thickness vs. the reflectance target across 400–700 nm for the 1.45/2.30 index pair. **(c)** Total thickness vs. the material figure of merit $(n_H-n_L)^2/(n_H+n_L)$ for twelve index pairs at 400–700 nm and $R = 99.9\%$. In (c), the bound is shown as a point for each of the twelve pairs, with a thin line as a guide for the eye.

## 4. Oblique incidence, dispersion, and loss

All of the analysis and examples in Figs. 1–3 have been at normal incidence. If a dielectric mirror is specified over a range of angles, a reasonable extension of our problem statement would be that the reflectance not fall below the target at every angle in the range and for all polarizations. At angle of incidence $\theta$, the transverse wavenumber scales with frequency, so the phase thickness of every layer remains proportional to $1/\lambda_0$ and the tilted stack is still a lossless, non-dispersive one-dimensional problem. The argument behind Eq. (1) therefore applies at each angle, with the wave admittance taking the role of the index, as in the oblique-incidence version of the sum rule for periodic screens [18]. For a layered profile, this gives:

$$\int_0^\infty \ln \frac{1}{|t(\lambda_0)|}\, \mathrm{d}\lambda_0 = \frac{\pi^2}{2Y_0} \int_0^D \frac{p(z)}{Y(z)} (Y(z) - Y_0)^2\, \mathrm{d}z, \tag{16}$$

where $p = \sqrt{n^2 - n_0^2 \sin^2\theta}$ is the "longitudinal index" that sets each layer's phase, $Y$ is the wave admittance ($Y = p$ for s-polarization and $Y = n^2/p$ for p-polarization) [4], and $Y_0$ is the admittance of the incident medium. Here we assume that $p(z)$ is real throughout the profile, meaning that no layer is beyond its critical angle ($n(z) > n_0 \sin\theta$ everywhere); this holds automatically for incidence from air, since every film index exceeds 1. At normal incidence, $p = Y = n$, and Eq. (16) reduces to Eq. (1).

Increasing the angle of incidence from the normal reduces p-polarized reflection as each interface approaches its Brewster condition. More generally, the right-hand side of Eq. (16) evaluated for p-polarization is never larger than its s-polarization value, for any profile at any angle. The smaller p-polarization budget therefore yields the larger thickness bound, and a design meeting the specification for both polarizations must in particular meet it for p-polarization, so that larger bound is the one that applies. Repeating the steps that led from Eq. (1) to Eq. (13), now starting from Eq. (16), with the admittance taking the place of the index in the matching and interface steps, can give a p-polarization thickness bound at each angle $\theta$. We can take the largest of these as an overall oblique-incidence bound; requiring a single stack to reach the target reflectance at every angle simultaneously can only raise the minimum thickness further, so the largest per-angle value remains a valid lower bound.

To test the per-angle bound, we used the 400–700 nm, $R = 99.9\%$ target specification at every angle from 0 to $\theta_{\text{max}}$, in air on both sides for simplicity, for $\theta_{\text{max}} = 0$, 15, 30, 45, and 60 degrees. We performed 20 optimizer runs for the 1.45/2.30 index pair. Fig. 4(a) shows the thinnest designs found, which are again roughly 2× the bound.

Figure 4(b) shows why the angular requirement results in increasing thickness. As the angle increases, the reflection band blue-shifts, and narrow features in the spectrum appear inside the target band and deepen with angle. The designs optimized over an angular range must cover a wider band than an equivalent normal-incidence design, wide enough that the shift with angle never uncovers the target band, and the resulting spectra should be free of narrow and particularly deep features over the specified range [Fig. 4(c)].

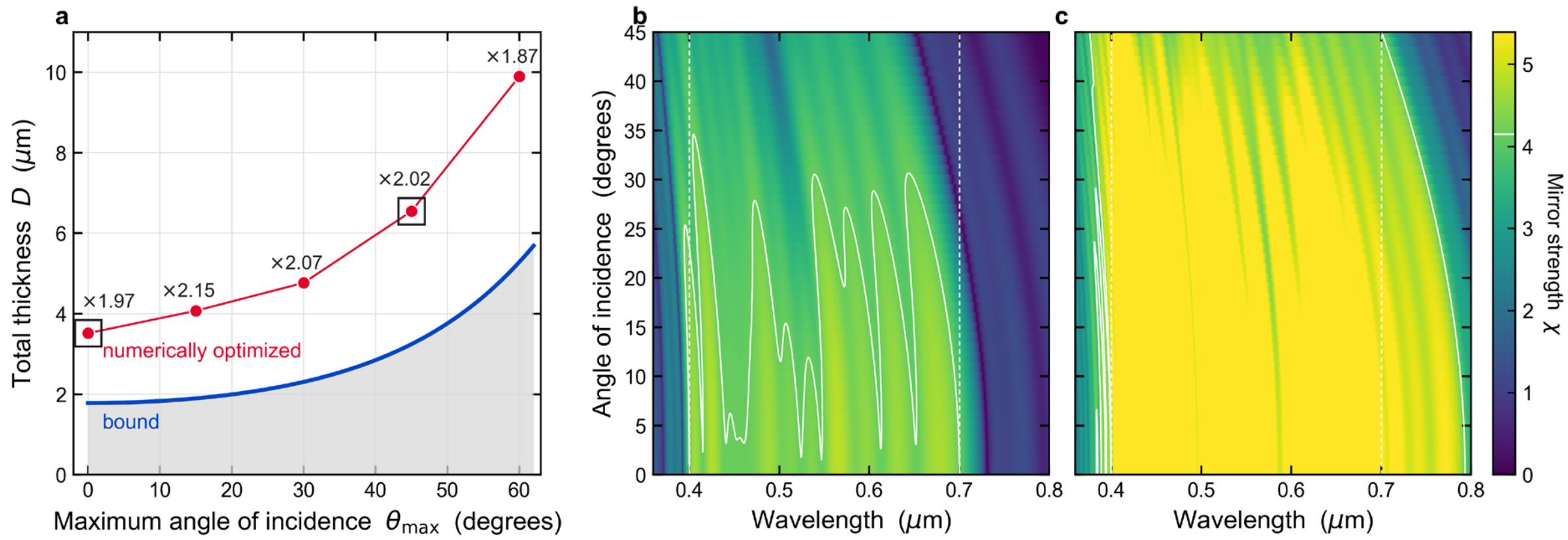

**Figure 4.** Dielectric mirrors designed to work over a range of angles from normal incidence to $\theta_{\max}$. The optimization was performed for the 1.45/2.30 pair in air, targeting at least $R = 99.9\%$ at every wavelength from 400 to 700 nm and every angle of incidence from 0 to $\theta_{\max}$, for both polarizations. **(a)** The bound and optimized total thickness vs. $\theta_{\max}$, where each optimized design meets the reflectance specifications over all angles from 0 to $\theta_{\max}$. The boxed points mark the two designs whose mirror-strength maps are shown in (b) and (c). **(b)** The p-polarized mirror strength of the $D = 3.52$ μm normal-incidence design as a function of wavelength and angle. Narrow transmission windows ("holes") open inside the target band. The white contour marks $\chi_{\text{target}}$, and the vertical dashed lines in (b) and (c) mark the 400–700 nm target band. **(c)** The p-polarized mirror strength for a design optimized for angles up to 45°, which meets the target specification throughout both the wavelength and angular range, and has $D = 6.55$ μm.

To confirm that the results are valid for dispersive refractive indices and asymmetric superstrates and substrates, we re-optimized the thin-film assemblies using dispersive refractive indices of materials widely used in thin-film coatings (though we still assume lossless films), with the incident medium being air and the substrate being BK7 glass. The $SiO_2$ and $Nb_2O_5$ indices were fitted to tabulated measurements of magnetron-sputtered films [19, 20], and for $Ta_2O_5$ we used the Cauchy formula fitted to reactive-magnetron-sputtered films [21]. Twenty-one optimizer runs covered seven target specifications for normal incidence, spanning wavelength bands from 400 to 1100 nm and reflectance targets from 99% to 99.99%, using the same optimizer as in Fig. 3 but with the wavelength-dependent indices. As with constant indices, the thinnest resulting designs that meet the specification are close to 2× the bound, as calculated for the smallest and largest values of the wavelength-dependent indices [Fig. 5(b)].

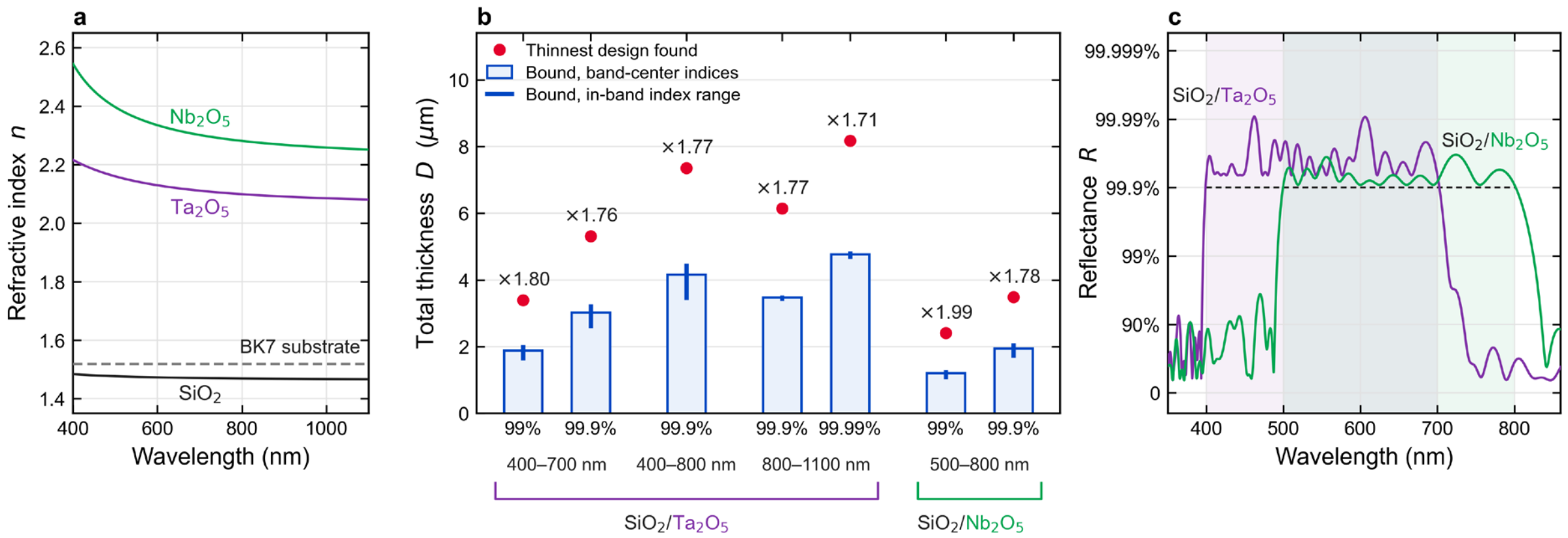

**Figure 5.** Dielectric mirrors optimized with measured dispersive indices for $SiO_2$, $Ta_2O_5$, and $Nb_2O_5$, with the incident medium being air ($n_0 = 1$) and the substrate being BK7 glass ($n_{\text{sub}} = 1.52$, assumed nondispersive), for normal incidence. **(a)** The refractive indices of $SiO_2$ and $Nb_2O_5$ were fitted to tabulated measurements of magnetron-sputtered films [19, 20], the index $Ta_2O_5$ is from the Cauchy formula for reactive-magnetron-sputtered films [21], and the BK7 substrate index is taken to be a constant 1.52. **(b)** The thinnest numerically optimized designs found for seven target specifications (red symbols). The bars mark the bound of Eq. (13), using the constituent refractive indices at the center of each wavelength band; the vertical whiskers span the bound obtained using the smallest and the largest values of each wavelength-dependent index within the wavelength band. **(c)** Reflectance spectra of two of the numerically optimized designs shown in (b), computed with the wavelength-dependent indices: the 68-layer, $D = 5.31$ μm $SiO_2/Ta_2O_5$ design targeting $R \geq 99.9\%$ over 400–700 nm, and the 44-layer, $D = 3.48$ μm $SiO_2/Nb_2O_5$ design targeting $R \geq 99.9\%$ over 500–800 nm.

We finally note that all of our calculations and optimizations above assumed lossless films. For reflectance targets up to $R = 99.9\%$, the best films of the materials considered here have low enough absorption for this idealization to be safe [22, 23]; for the highest targets, such as 99.99% and above, $1 - R$ becomes comparable to the absorptance of even the best available films [1, 23], and absorption has to be treated explicitly.

## 5. Summary

We identified a thickness bound for dielectric mirrors, which originates in a causality-based sum rule previously derived in one-dimensional acoustics. The sum rule states that, for a lossless, non-dispersive index profile, the all-wavelength integral of logarithmic transmittance is determined by how much material of each index is present, and rearranging the layers merely redistributes that integral in wavelength. We converted this observation into a lower bound on the physical thickness required for a specified wavelength span and worst-case reflectance of a dielectric mirror. In our numerical tests, we consistently found optimized designs at roughly twice the thickness bound, largely because reflection outside the target band and reflectance above the target within the band consume part of the fixed reflectance "budget" from the sum rule. Although a thin-film stack can readily be optimized directly, our analytical bound provides physical intuition and useful design rules, including a figure of merit for material selection. For dielectric mirrors with extreme requirements, such as very broad wavelength bands, where both thickness and layer count can become large, our bound can be a reference for judging whether an optimizer in a complex landscape with many degrees of freedom is approaching the minimum possible thickness.

## Acknowledgement

Most of the numerical optimizations were performed using the UW-Madison Center for High Throughput Computing (CHTC). Thanks to Tanuj Kumar and Jin-Woo Cho for critical reading of the manuscript.

## Disclosure of AI use, and discussion

I thought of this problem while working on some related thin-film optics problems and having in mind the Rozanov bound for absorbers [6]. Inspired by recent AI-aided results in mathematics, I decided to aim three current frontier LLM-based AI tools at the task of identifying a thickness bound for dielectric mirrors: Claude Fable 5, Opus 5, and OpenAI GPT-5.6 Sol. I threw the attempted solutions back and forth between the tools, and what emerged was the sum rule of Eq. (1), which was converted from the acoustics result by Norris [7]. These tools also helped me come up with the actual bounds of Eqns. (13-15). While this was not a fully automated process, and I did nudge the tools in various directions, I have to say that the bound was found mostly by the AI tools. After checking the results with numerical optimization, and excited by getting an interesting result so quickly, I tried to have the AI tools draft a manuscript, the result of which was utterly unreadable and contained errors. Over the next month I checked and refined the arguments and logic, fixed the figures and made new ones, ran more numerical checks, and wrote the manuscript in an iterative process with assistance from the AI tools. This has been a new experience for me.